\documentclass[twocolumn,aps,floatfix,superscriptaddress]{revtex4}
\pdfoutput=1 
\usepackage{amsmath,amssymb,eucal,graphicx,float,epstopdf}

\usepackage[colorlinks=true, urlcolor=blue, anchorcolor=blue, citecolor=blue,filecolor=blue,linkcolor=blue,menucolor=blue]{hyperref}

\begin{document}
\title{Large Deviations in One-Dimensional Random Sequential Adsorption}

\author{P.~L.~Krapivsky}
\affiliation{Department of Physics, Boston University, Boston, Massachusetts 02215, USA}

\begin{abstract} 
In random sequential adsorption (RSA), objects are deposited randomly, irreversibly, and sequentially; attempts leading to an overlap with previously deposited objects are discarded. The process continues until the system reaches a jammed state when no further additions are possible.  We analyze a class of lattice RSA models in which landing on an empty site in a segment is allowed when at least $b$ neighboring sites on the left and the right are unoccupied. For the minimal model ($b=1$), we compute the full counting statistics of the occupation number. We reduce the determination of the full counting statistics to a Riccati equation that appears analytically solvable only when $b=1$. We develop a perturbation procedure which, in principle, allows one to determine cumulants consecutively, and we compute the variance of the occupation number for all $b$. 
\end{abstract}

\maketitle

\section{Introduction}
\label{sec:Intro}

Random sequential adsorption (RSA) is a toy model mimicking the irreversible deposition of suspended particles onto substrates. The RSA model postulates that the deposition events are random: If the new particle is sufficiently far away from already deposited ones, it sticks to the substrate; otherwise, the deposition event is discarded.  In the two-dimensional setting, the RSA models have been applied to modeling chemisorption on single-crystal surfaces and adsorption in colloidal systems \cite{Evans,TT,Tor,Adam,Newby13}. There are also applications in nanotechnology, see \cite{Elimelech02,Kuznar03,Floro06,Yaish13}. The RSA models have been also used in high dimensions, e.g., in the context of packing problems  \cite{TS10}. 

The first RSA model with adsorption of dimers was introduced by Flory for the description of reactions along a long polymer chain \cite{F39}. Another beautiful RSA model with adsorption on a continuous one-dimensional line was introduced by R\'enyi \cite{R58} as a toy model of car parking. The RSA type of models has been also used in several other one-dimensional settings, e.g., in modeling polymer translocation \cite{DCA07,KM10} and describing zero-temperature dynamics of Ising chains \cite{Privman92,PK94,JML02}. 

The RSA models mimic the generation Rydberg excitations. In experiments, a laser excites ultra-cold atoms on a lattice from the ground state to a Rydberg state (see e.g. \cite{Exp:00,Exp:05,Exp:12,Exp:13,Exp:14,Rydberg:FCS}). Interactions between Rydberg atoms cause the blockage forbidding the excitation of atoms sufficiently close to a Rydberg atom. When the radiative decay of the Rydberg atoms can be ignored, this RSA model mimics certain features of the excitation process \cite{Dutch14}, although it disregards features like  the non-ergodic quantum dynamics of Rydberg-blockaded chains (see \cite{Lukin17,Abanin17,Lukin18,Chris18} and references therein). 

We investigate a model in which particles are absorbed randomly and irreversibly onto an interval with $L$ sites. We assume that each particle attaches to a single site. The blockage effect is modeled by the requirement that adjacent particles are separated by at least $b$ vacant sites. In experimental realizations with Rydberg atoms where excitation plays the role of deposition, these systems typically occupy $L \sim 50$ lattice sites. For such relatively small systems, fluctuations are significant, and they have been probed experimentally. 

Strongly sub-Poissonian statistics has been observed in experiments  \cite{Exp:05,Exp:12,Exp:13,Exp:14,Rydberg:FCS} with Rydberg atoms, and it has been attributed to the blockage effect. The RSA process provides an approximate treatment of some features of Rydberg gases (see \cite{Dutch14} for more complicated modeling relying on classical stochastic processes), and it captures the strongly sub-Poissonian statistics as we demonstrate in this paper. The most basic deviation from the Poisson statistics is quantified by the so-called Mandel $Q$ parameter varying in the range $-1<Q<\infty$, and vanishes for the Poisson process. Below we compute $Q_b$ for all $b$; e.g., $Q_1\approx -0.957635\ldots$ and $Q_2\approx -0.9518$ demonstrating a strongly sub-Poissonian behavior.

The average occupation number in a model mathematically isomorphic to the minimal model was computed by Flory \cite{F39}. 
We outline his computation in Sec.~\ref{sec:MM} and then extend it to determine the behaviors near the boundary, e.g., the probabilities that the leftmost site, and those second and third from the left are occupied. We also determine the total number of jammed states. In Sec.~\ref{sec:LD}, we compute the full counting statistics of the occupation number for the minimal model. In the general case when the  blockage radius is arbitrary (Sec.~\ref{sec:LMb}), the average occupation number is also computable; the generating function encoding all the cumulants satisfies a Riccati equation which appears unsolvable for any $b\geq 2$. Cumulants can, in principle, be extracted one by one via increasingly more cumbersome perturbative calculations, so we only determine the variance (the computations are relegated to the appendix). The extreme probabilities of maximally sparse and maximally dense jammed configurations are tractable for arbitrary $b\geq 1$.

\section{The Minimal Model: Average Properties}
\label{sec:MM}

The exclusion interaction between particles is modeled by requiring that the two sites surrounding an occupied site are empty. Thus there is the blockade effect as each absorbed particle forbids future adsorption into the nearest-neighbor sites. 

Eventually, the system reaches a jammed state. For instance,
\begin{equation}
\label{jammed}
\bullet\, \circ\,\bullet\,\circ\,\circ\,\bullet\,\circ\,\circ\,\bullet\,\circ\,\bullet\,\circ\,\circ\,\bullet\,\circ\,\bullet\,\circ\,\bullet\,\circ\,\,\bullet
\end{equation}
is a jammed configuration with $N=9$ occupied sites on the interval of length $L=19$. 

Jammed states vary from realization to realization, yet global characteristics of jammed states become deterministic in the thermodynamic $L\to\infty$ limit. For instance, the fraction of short segments $~\bullet\,\circ\,\bullet~$ is $1-3e^{-2}$, while the fraction of the long segments $~\bullet\,\circ\,\circ\,\bullet~$ is $3e^{-2}$. Since we study fluctuations, so we must analyze finite systems. We consider open intervals that are relevant in the context of Rydberg atoms; the extension to rings is straightforward.  

\subsection{The average number of occupied sites}

Here we outline the computation of the average number $A_L=\langle N\rangle$ of occupied sites on an open interval of length $L$. We employ an approach essentially developed by Flory \cite{F39}, see also a textbook \cite{book}; our analysis (Sec.~\ref{sec:LD}) of the cumulant generating function will proceed along similar lines.  

For small $L$, one can compute $A_L=\langle N\rangle$ by hand. For instance, $N=1$ when $L=1$ and 2; for $L=4$, one gets $N=2$. For $L= 3$ and $L\geq 5$, the total number $N$ of absorbed particles varies from realization to realization, e.g. $N=1$ (probability $1/3$) or $N=2$ (probability $2/3$) when $L=3$. Thus
\begin{equation}
A_1=1,\quad A_2=1, \quad A_3 = \frac{5}{3}, \quad A_4 = 2
\end{equation}

To compute $A_L$ for arbitrary $L$, suppose $k$ is the landing site of the first particle. The intervals of lengths $k-2$ and $L-k-1$ on the left and right of the first deposited particle are subsequently filled {\em independently}. This key feature makes the one-dimensional situation tractable. For our model, we arrive at the recurrence
\begin{equation}
\label{AL:rec}
A_L = \frac{1}{L}\sum_{k=1}^L \big(A_{k-2}+1+A_{L-k-1}\big)
\end{equation}
that is applicable for all $L\geq 1$ if we set $A_{-1}=A_0=0$. Using the generating function 
\begin{equation}
\label{Ax:def}
A(x) = \sum_{L\geq 1} A_{L}\, x^L
\end{equation}
we convert the recurrence \eqref{AL:rec} into a differential equation
\begin{equation}
\frac{dA(x)}{dx} = \frac{2x}{1-x}\,A(x)+\frac{1}{(1-x)^2}
\end{equation}
Solving this linear inhomogeneous differential equation subject to the initial condition $A(0)=0$, we obtain
\begin{equation}
\label{Ax}
A(x) = \frac{1-e^{-2x}}{2(1-x)^2}
\end{equation}
from which
\begin{equation}
\label{AL:sol}
A_L = \sum_{k=0}^{L-1} \frac{(-2)^k}{(k+1)!}\,(L-k)
\end{equation}
Using \eqref{AL:sol} one can extract the large $L$ behavior:
\begin{equation}
\label{AL:asymp}
A_L = \frac{1-e^{-2}}{2}\,(L+ 3) - 1 + \frac{(-2)^{L+1}}{(L+2)!}+\ldots
\end{equation}

\subsection{Behavior near the boundary}

The average jamming density is $\rho_L =A_L/L$, and in the $L\to\infty$ limit the jamming density becomes
\begin{equation}
\label{prob:bulk}
\rho_\text{jam} =\rho_\infty = \frac{1-e^{-2}}{2}=0.432332358\ldots
\end{equation}
When $L\gg 1$, the probability that a site in the bulk is occupied is very close to \eqref{prob:bulk}. Near the boundary the densities are different. 

Let us first compute the probability $p_L$ that the leftmost site is occupied. For small $L$, one can compute $p_L$ by hand to yield 
\begin{equation}
p_1=1,\quad p_2=\frac{1}{2}, \quad p_3 = \frac{2}{3}, \quad p_4 = \frac{5}{8}
\end{equation}
etc. Generally, the probability $p_L$ satisfies the recurrence
\begin{equation}
\label{pL:rec}
p_L =  \frac{1}{L} + \frac{1}{L}\sum_{k=1}^{L-2} p_{k}
\end{equation}
Indeed, with probability $L^{-1}$, the first landing site is the leftmost site; this explains the first term on the right-hand side of \eqref{pL:rec}. The leftmost site may also get occupied if the first landing site is $k+1$ with $k=1,\ldots,L-2$. Each such event happens with probability $L^{-1}$, and the leftmost site in the segment of length $k$ could be occupied with probability $p_k$. This explains the sum on the right-hand side of \eqref{pL:rec}. 

Using the generating function 
\begin{equation}
\label{Px:def}
P(x) = \sum_{L\geq 1} p_{L}\, x^L
\end{equation}
we convert the recurrence \eqref{pL:rec} into a differential equation
\begin{equation}
\frac{dP(x)}{dx} = \frac{x}{1-x}\,P(x)+\frac{1}{1-x}
\end{equation}
which is solved to yield
\begin{equation}
\label{Px:sol}
P(x) = \frac{1-e^{-x}}{1-x}
\end{equation}
Expanding $P(x)$ we obtain
\begin{equation}
\label{pL:sol}
p_L = \sum_{k=1}^L \frac{(-1)^{k-1}}{k!}
\end{equation}
In the $L\to\infty$ limit, i.e. for the semi-infinite lattice, the left-most site is occupied with probability
\begin{equation}
\pi_1 =p_\infty = 1-e^{-1} =0.6321205588\ldots
\end{equation}
which substantially exceeds the probability \eqref{prob:bulk} that a bulk site is occupied. 

The probability $q_L$ that the second site, viz. the site adjacent to the left-most site is occupied, is dual to the probability $p_L$: 
\begin{equation}
\label{qL:sol}
q_L = 1 - p_L = \sum_{k=0}^L \frac{(-1)^{k}}{k!}
\end{equation}
for $L\geq 2$. In particular
\begin{equation}
\pi_2 =q_\infty = e^{-1} =0.36787944117\ldots
\end{equation}

The probability $r_L$ that the third site is occupied satisfies the recurrence
\begin{equation}
\label{rL:rec}
r_L =  \frac{1}{L} + \frac{1}{L}\sum_{k=3}^{L-2} r_{k} + \frac{1}{L}\,p_{L-2}
\end{equation}
which is established similarly to the recurrence \eqref{pL:rec}. 
Using the generating function 
\begin{equation}
\label{Rx:def}
R(x) = \sum_{L\geq 3} r_{L}\, x^L
\end{equation}
we convert the recurrence \eqref{rL:rec} into a differential equation
\begin{equation}
\label{Rx:eq}
\frac{dR(x)}{dx} = \frac{x}{1-x}\,R(x)+\frac{x^2}{1-x}+xP(x)
\end{equation}
with $P(x)$ given by \eqref{Px:sol}. Solving \eqref{Rx:eq} subject to $R(0)=0$ we obtain
\begin{equation}
\label{Rx:sol}
R(x) = \frac{1-x+x^2-\left(1+\frac{x^2}{2}\right)e^{-x}}{1-x}
\end{equation}
from which 
\begin{equation}
\label{rL:sol}
r_L = 1 -  \frac{1}{2}\sum_{k=0}^{L-2} \frac{(-1)^{k}}{k!}  - \sum_{k=0}^L \frac{(-1)^{k}}{k!}
\end{equation}
In particular,
\begin{equation}
\pi_3 =r_\infty = 1-\tfrac{3}{2}e^{-1} = 0.4481808382\ldots
\end{equation}

A longer calculation gives
\begin{equation}
\pi_4= \tfrac{37}{4}e^{-1} - 3 = 0.4028848308\ldots
\end{equation}

These results hint that
\begin{equation*}
\pi_{2n} = A_n e^{-1} - B_n, \qquad \pi_{2n-1} = C_n - D_n e^{-1} 
\end{equation*}
with some positive rational $A_n, B_n, C_n, D_n$. Recalling that $\pi_\infty=\rho_\text{jam}$, one gets
\begin{equation*}
\begin{split}
&\lim_{n\to\infty} \left[1+2B_n-2A_n e^{-1}\right] = e^{-2} \\
&\lim_{n\to\infty} \left[1-2C_n + 2D_n e^{-1}\right] = e^{-2}
\end{split}
\end{equation*}

\subsection{The total number of jammed states}

The total number of jammed states $J_L$ satisfies the recurrence
\begin{equation}
\label{JL-rec}
J_L = J_{L-2}+J_{L-3}
\end{equation}
Indeed, the jammed states can be divided into the complementary sets with the occupied and empty leftmost site. If in a jammed state the leftmost site is occupied, the second site must be empty; the total number of such jammed states is equal to $J_{L-2}$. If the leftmost site in a jammed state is empty, the second site must be occupied and the third must be empty; the total number of such jammed states is equal to $J_{L-3}$.

The solution of the recurrence \eqref{JL-rec} subject to the boundary conditions $J_0=J_1=1$ and $J_2=2$ is encapsulated in the generating function
\begin{equation}
\sum_{L\geq 0}J_L x^L = \frac{1+x+x^2}{1-x^2-x^3}
\end{equation}
In particular, the asymptotic behavior is
\begin{equation}
J_L \simeq \frac{(1+r)^2}{2r+3}\,r^L\quad\text{as}\quad L\to\infty
\end{equation}
where $r$ is the real root of the cubic equation $r^3=r+1$:
\begin{eqnarray*}
r  &=&\sqrt[3]{1+\sqrt[3]{1+\sqrt[3]{1+\ldots}}} \\
   &=& \frac{\sqrt[3]{108+12\sqrt{69}} + \sqrt[3]{108-12\sqrt{69}}}{6} \\
   &=&1.3247179572\ldots
\end{eqnarray*}
known as the plastic number. The Padovan sequence \eqref{JL-rec} is a cousin of the Fibonacci sequence, and it also arises in several applications (see, e.g., \cite{Zagier,JML}).

\section{The Minimal Model: Full counting statistics}
\label{sec:LD}

One-dimensional RSA models are often tractable (see \cite{Evans,TT,book} for a review) and basic features like the jamming density have been probed analytically. For the minimal model with $b=1$, we have shown how to compute the jamming density, the behavior near the boundary, and the total number of jammed states  (Sec.~\ref{sec:MM}). In this section we analyze fluctuations.

\subsection{Full counting statistics}

Consider an interval of length $L$. The RSA procedure brings the system into a jammed state. The total number $N$ of absorbed particles fluctuates from realization to realization.  Here we compute the cumulant generating function that encodes all the cumulants of $N$. Thus we want to compute the average 
\begin{equation}
\label{F:def}
F(\lambda, L) \equiv \langle e^{\lambda N}\rangle = \sum_N e^{\lambda N} P(N,L)
\end{equation}
where $P(N,L)$ is the probability to have $N$ particles in a jammed state. The standard relation
\begin{equation}
\ln \langle e^{\lambda N}\rangle = \sum_{n\geq 1} \frac{\lambda^n}{n!}\,  \langle N^n\rangle_c
\end{equation}
then gives all the cumulants: the average $\langle N\rangle_c = \langle N\rangle$, the variance $\langle N^2\rangle_c = \langle N^2\rangle - \langle N\rangle^2$, etc. 

The function $F(\lambda, L) \equiv \langle e^{\lambda N}\rangle$ grows exponentially with $L$. The cumulant generating function 
\begin{equation}
\label{U:def}
U(\lambda) = \lim_{L\to\infty} L^{-1} \ln F(\lambda, L)
\end{equation}
encapsulates all cumulants: 
\begin{equation}
U(\lambda)=\sum_{n\geq 1} \frac{\lambda^n}{n!}\, U_n, \qquad \langle N^n\rangle_c = L U_n
\end{equation}

The cumulant generating function satisfies 
\begin{equation}
\label{FL:rec}
F(\lambda, L) = \frac{e^\lambda}{L}\sum_{k=1}^L F(\lambda, k-2) F(\lambda, L-k-1)
\end{equation}
To derive this recurrence, suppose $k$ is the first landing site. The intervals on the left and right of site $k$ are filled independently. (The one-dimensional nature of the problem is crucial for this property.)  If $N_-$ and $N_+$ are the final occupation numbers of the left and right intervals, the total occupation number is $N= N_- + 1 +N_+$, so $e^{\lambda N}=e^{\lambda}e^{\lambda N_-}e^{\lambda N_+}$. Performing the averaging, summing over all $k$, and taking into account that site $k$ is chosen with probability $L^{-1}$, we obtain Eq.~\eqref{FL:rec}.

\begin{figure}[ht]
\begin{center}
\includegraphics[width=0.4\textwidth]{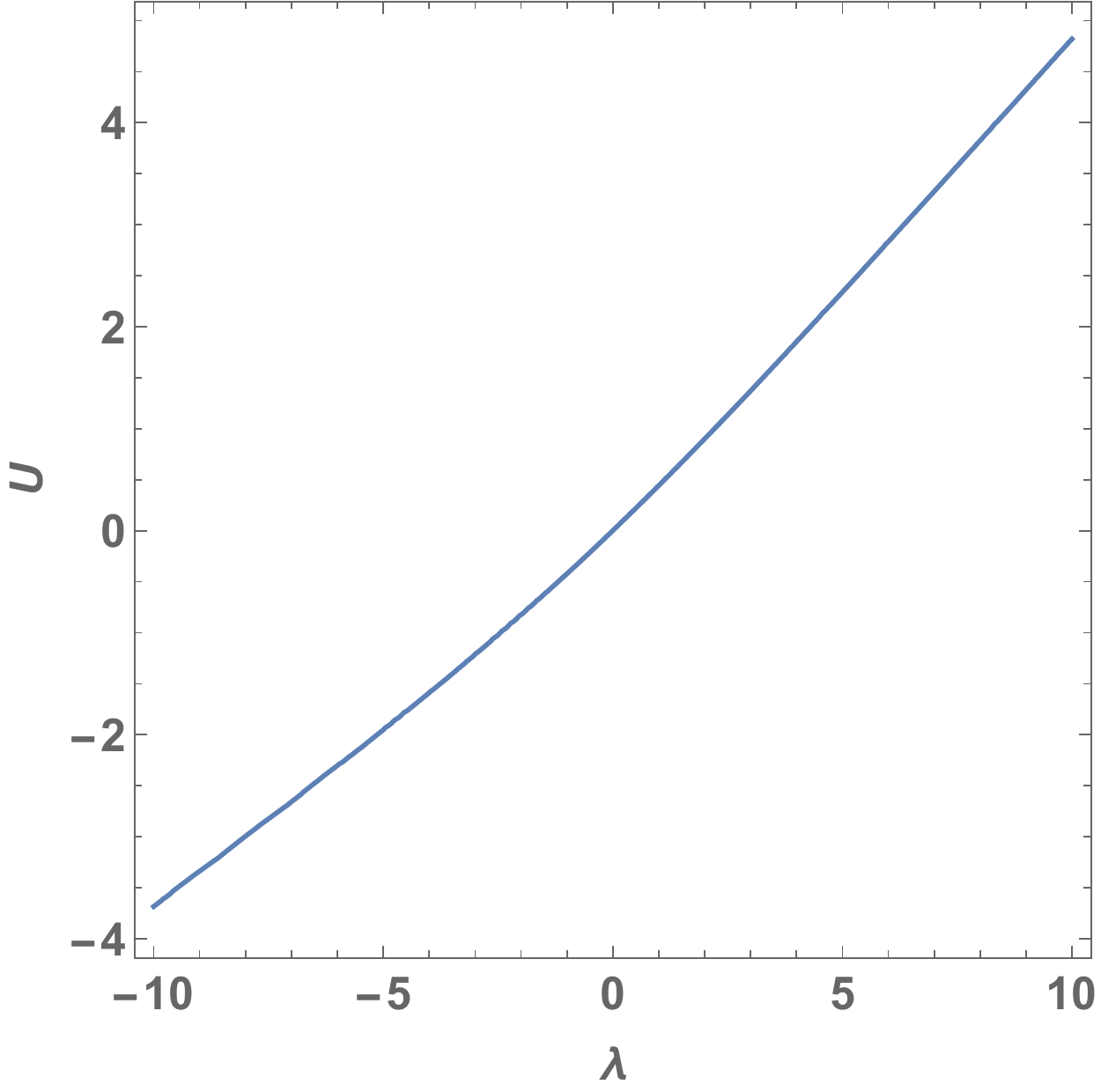}
\caption{The plot of the cumulant generating function $U(\lambda)$. The expansion of $U(\lambda)$ at $\lambda=0$ yields the cumulants.}
\label{fig:UL}
  \end{center}
\end{figure}

The recurrence \eqref{FL:rec} applies for all $L\geq 1$ if we set
\begin{equation}
\label{IC:FF}
F(\lambda, -1) = F(\lambda, 0) = 1
\end{equation}
We now introduce the generating function
\begin{equation}
\label{Phi:def}
\Phi(\lambda, x) = \sum_{L\geq 0}  F(\lambda, L)\, x^L
\end{equation}
To recast \eqref{FL:rec} into an equation for the generating function we multiply \eqref{FL:rec} by $Lx^{L-1}$ and sum over all $L\geq 1$. The left-hand side turns into
\begin{equation*}
\sum_{L\geq 1}  L F(\lambda, L)\, x^{L-1} = \frac{\partial \Phi(\lambda, x)}{\partial x}
\end{equation*}
The right-hand side becomes
\begin{equation*}
e^\lambda \sum_{L\geq 1} \sum_{k=1}^L F(\lambda, k-2) F(\lambda, L-k-1)\, x^{L-1}
\end{equation*}
and using the boundary conditions \eqref{IC:FF} we simplify the double sum to
\begin{equation*}
 x^2 \sum_{m\geq -1}  F(\lambda, m)\, x^m \sum_{n\geq -1}  F(\lambda, n)\, x^n = \left[1+x\Phi(\lambda, x)\right]^2
\end{equation*}
and recast the recurrence \eqref{FL:rec} into a Riccati equation
\begin{equation}
\label{Phi:eq}
\frac{d \Phi}{d x}  = e^\lambda \left[1+x\Phi\right]^2
\end{equation}
which turns out to be solvable. The solution of Eq.~\eqref{Phi:eq} subject to the initial condition $\Phi(\lambda, 0)=1$ reads 
\begin{equation}
\label{FLx}
\Phi(\lambda, x) = \frac{1+\Lambda \tanh(\Lambda x)}{1-x+(\Lambda^{-1}-\Lambda  x)\tanh(\Lambda x)}
\end{equation}
where $\Lambda\equiv e^{\lambda/2}$. The generating function $\Phi(\lambda, x)$ has a simple pole at $x=y=y(\lambda)$. Using \eqref{FLx}  we find that the pole is implicitly determined by
\begin{equation}
\label{y:def}
\frac{\tanh(\Lambda y)}{\Lambda} = \frac{y-1}{1-\Lambda^2 y}\,, \qquad \Lambda = e^{\lambda/2}
\end{equation}
The cumulant generating function 
\begin{equation}
\label{U:implicit}
U(\lambda) = -\ln y(\lambda)
\end{equation}
is plotted in Fig.~\ref{fig:UL}. 

We already know the  dominant exponential factor in $F(\lambda, L) \propto e^{LU(\lambda)}$.  The exact result \eqref{FLx} allows one to find a pre-exponential factor. A straightforward calculation gives the residue of the simple pole of $\Phi(\lambda, x)$. We find $\Phi(\lambda, x) = (\Lambda y)^{-2}(y-x)^{-1} + O(1)$ when $x\to y$, from which we deduce 
\begin{equation}
\label{FU}
F(\lambda, L) \simeq (\Lambda^2 y^3)^{-1}\, e^{LU(\lambda)}
\end{equation}
when $L\gg 1$. 

The cumulant generating function is implicitly defined through \eqref{y:def}--\eqref{U:implicit}. Expanding $U(\lambda)$, we recover the already known value of the first cumulant: $\langle N\rangle = L \rho_\text{jam}$. In principle, one can determine an arbitrary cumulant. We haven't succeeded in finding an explicit general formula. Here we merely list a few cumulants that are extracted by expanding $U(\lambda)$ implicitly given by \eqref{y:def}--\eqref{U:implicit}: 
\begin{equation}
\begin{split}
\frac{\langle N^2\rangle_c}{L} & = \frac{1}{e^4}\\
\frac{\langle N^3\rangle_c}{L} & = \frac{e^4-61}{16\,e^6}\\
\frac{\langle N^4\rangle_c}{L} & =  \frac{43-e^4}{2\,e^8}\\
\frac{\langle N^5\rangle_c}{L} & =  \frac{305e^4 - 10283 - 2e^8}{64\,e^{10}}\\
\frac{\langle N^6\rangle_c}{L} & =  \frac{47871 - 1720 e^4 + 17 e^8}{32\,e^{12}}\\
\frac{\langle N^7\rangle_c}{L} & =  \frac{359905 e^4 + 17 e^{12} - 8540689 - 4697  e^8}{512\,e^{14}}\\
\frac{\langle N^8\rangle_c}{L} & =  \frac{6935126 - 335097 e^4 + 5418 e^8 - 31 e^{12}}{32\,e^{16}}
\end{split}
\end{equation}
The so-called  Mandel $Q$ parameter \cite{Mandel} defined via
\begin{equation}
\label{Q:def}
Q=\frac{\langle N^2\rangle_c}{\langle N\rangle}-1
\end{equation}
is a basic measure characterizing the deviation from Poissonian statistics. The values  $-1\leq Q<\infty$ are permissible, for the Poisson statistics $Q=0$ and the range $-1\leq Q<0$ is sub-Poissonian. The numerical value
\begin{equation}
\label{Q:RSA}
Q=\frac{2}{e^2(e^2-1)}-1 = -0.95763528097389\ldots
\end{equation}
indicates that the statistics is strongly sub-Poissonian in the present case. 

The ratios $\langle N^n\rangle_c/\langle N\rangle$ of cumulants to the average are known as Fano factors \cite{Fano}.
Here are a few Fano factors 
\begin{equation*}
\begin{split}
\frac{\langle N^3\rangle_c}{\langle N\rangle} & = \frac{e^4-61}{8\,e^4(e^2-1)}= -0.0022940394167\ldots\\
\frac{\langle N^4\rangle_c}{\langle N\rangle} & =  \frac{43-e^4}{e^6(e^2-1)}= -0.00449971626\ldots\\
\frac{\langle N^5\rangle_c}{\langle N\rangle} & =  \frac{305e^4 - 10283 - 2e^8}{16\,e^{10}(e^2-1)}=0.00009049346\ldots\\
\frac{\langle N^6\rangle_c}{\langle N\rangle} & =  \frac{47871 - 1720 e^4 + 17 e^8}{16\,e^{12}(e^2-1)}=0.0002787944\ldots
\end{split}
\end{equation*}
For the Poisson distribution, all Fano factors are equal to unity. Thus Fano factors further illustrate a substantial deviation from the Poisson statistics. 

\subsection{Extremal jammed states}
\label{subsec:Extremal}

Extremal jammed states are the states with the largest and the smallest number of occupied sites. Let us extract the probabilities of such states from the behavior of the cumulant generating function in the $\lambda\to \pm \infty$ limits. An asymptotic analysis of \eqref{y:def} yields (see also Fig.~\ref{fig:UL})
\begin{equation}
\label{U:asympt}
U(\lambda) = 
\begin{cases}
\frac{1}{3}\lambda - \frac{1}{3}\ln 3  & \lambda\to -\infty\\
\frac{1}{2}\lambda - \ln u                  & \lambda\to \infty
\end{cases}
\end{equation}
We have dropped the terms vanishing in the $\lambda\to \pm \infty$ limits and denoted by $u$ the positive root of the equation $u\,\tanh(u)=1$;  numerically $u=1.19967864\ldots$. Combining \eqref{FU} and \eqref{U:asympt} we get
\begin{equation}
\label{F:asympt}
F(\lambda, L) \sim 
\begin{cases}
3^{-L/3}~e^{\lambda L/3}      & \lambda\to -\infty\\
u^{-L}~e^{\lambda L/2}          & \lambda\to \infty
\end{cases}
\end{equation}

To appreciate the first asymptotic in \eqref{F:asympt} we note that in the $\lambda\to - \infty$ limit the dominant contribution to the sum in Eq.~\eqref{F:def} is provided by the jammed state with the smallest number of occupied sites. This jammed state 
\begin{subequations}
\begin{equation}
\label{Z3}
\ldots \circ\,\bullet\,\circ\,\circ\,\bullet\,\circ\,\circ\,\bullet\,\circ\,\circ\,\bullet\,\circ\,\circ\,\bullet\,\circ\,\circ\,\bullet\,\circ\,\circ\,\bullet\,\circ\ldots
\end{equation}
has $N_\text{min}=\lfloor (L+2)/3\rfloor$ particles explaining the dominant $e^{\lambda L/3}$ factor. The $3^{-L/3}$ pre-factor gives the probability to end up in such jammed state. 

In the $\lambda\to \infty$ limit the dominant contribution to the sum \eqref{F:def} is provided by the jammed state with the largest number of occupied sites:  
\begin{equation}
\label{Z2}
\ldots \circ\,\bullet\,\circ\,\bullet\,\circ\,\bullet\,\circ\,\bullet\,\circ\,\bullet\,\circ\,\bullet\,\circ\,\bullet\,\circ\,\bullet\,\circ\,\bullet\,\circ\ldots
\end{equation}
\end{subequations}
We have $N_\text{max}=\lfloor (L+1)/2\rfloor$ and this explains the $e^{\lambda L/2}$ factor in \eqref{F:asympt}. The $u^{-L}$ pre-factor is the probability to end up in the jammed state with the largest number of occupied sites. Thus the large deviation technique gives the asymptotic behaviors of the probability to reach the extremal jamming jammed states: 
\begin{subequations}
\begin{align}
\label{Prob:extreme-max}
& P(N_\text{max},L)\sim u^{-L} \\
\label{Prob:extreme-min}
& P(N_\text{min},L)\sim 3^{-L/3}
\end{align}
\end{subequations}

The probabilities $P(N_\text{max},L)$ and $P(N_\text{min},L)$ can be also computed exactly as follows. Let us start with maximally dense jammed states \eqref{Z2}. The maximally dense state with $N_\text{max}=n$ particles may occur when $L=2n-1$ or $L=2n$. The probability $\mu_n=P(n,2n-1)$ can be recurrently determined from 
\begin{equation}
\label{mu-n}
\mu_n = \frac{1}{2n-1}\sum_{k=0}^{n-1}\mu_k\mu_{n-k-1}
\end{equation}
Indeed, to ensure the maximal occupancy, the first landing must be to a site with an odd label, say $2k+1$. The interval of length $2k-1$ on the left (site $2k$ must remain unoccupied) and the interval of length $2(n-k-1)-1$ on the right are filled independently, and the requirement of the maximal occupancy gives the recurrence \eqref{mu-n}. 

The boundary conditions read 
\begin{equation}
\label{mu-n:BC}
\mu_0 = \mu_1 = 1
\end{equation}
Introducing the generating function 
\begin{equation}
\label{mu:GF}
\mu(x)=\sum_{k\geq 0}\mu_k x^{k}
\end{equation}
we convert the recurrence \eqref{mu-n} into a Riccati equation 
\begin{equation}
\label{mu:Ric}
2\,\frac{d\mu}{dx}=\frac{\mu-1}{x}+\mu^2
\end{equation}
which is solved to yield
\begin{equation}
\label{mu:GF-sol}
\mu(x) = \frac{1}{1-\sqrt{x}\, \tanh\sqrt{x}}
\end{equation}
This generating function has a simple pole $2/(u^2-x)$ at $x=u^2$, from which we extract the large $n$ asymptotic
\begin{equation}
\label{mu-asymp}
\mu_n = P(n,2n-1) \simeq \frac{2}{u^{2n+2}}
\end{equation}

The probability $\widehat{\mu}_n=P(n,2n)$ can be recurrently determined from 
\begin{equation}
\label{mu-n-hat}
\widehat{\mu}_n = \frac{1}{n}\sum_{k=0}^{n-1}\widehat{\mu}_k\,\mu_{n-k-1}
\end{equation}
which is derived similarly to \eqref{mu-n}. To appreciate the factor $n^{-1}$ in front of the sum we note that the probability of the first landing attempt is now $(2n)^{-1}$. The interval of even length may be either on the left or the right from the first landing site. This gives an extra factor of two.

The generating function
\begin{equation}
\label{mu-hat:GF}
\widehat{\mu}(x)=\sum_{k\geq 0}\mu_k x^{k}
\end{equation}
converts the recurrence \eqref{mu-n-hat} into a differential equation 
\begin{equation}
\label{mu-hat}
\frac{d\widehat{\mu}}{dx}=\widehat{\mu}\, \mu
\end{equation}
with known $\mu(x)$, see \eqref{mu:GF-sol}. Integrating \eqref{mu-hat} yields 
\begin{equation}
\label{mu-hat-sol}
\widehat{\mu}(x) = \frac{1}{\big(\cosh\sqrt{x}-\sqrt{x}\, \sinh\sqrt{x}\big)^2}
\end{equation}
This generating function behaves as
\begin{equation}
\widehat{\mu}(x) = \frac{4(1-u^{-2})}{(u^2-x)^2}+\frac{2(1-u^{-2})}{u^2(u^2-x)}+O(1)
\end{equation}
when $x\to u^2$, from which we extract the large $n$ asymptotic
\begin{equation}
\label{mu-asymp-hat}
\widehat{\mu}_n = P(n,2n) \simeq 2\left(1-u^{-2}\right)\frac{2n+3}{u^{2n}}
\end{equation}
The asymptotic behaviors \eqref{mu-asymp} and \eqref{mu-asymp-hat} are the more precise versions of the asymptotic \eqref{Prob:extreme-max} following from the cumulant generating function. 

Consider now the maximally sparse jammed states, say for intervals of length $L=3n$, so that $N_\text{min}=n$. The probabilities $\nu_n=P(n,3n)$ satisfy the recurrence
\begin{equation}
\label{nu-n}
\nu_n = \frac{1}{3n}\sum_{k=0}^{n-1}\nu_k\nu_{n-k-1}
\end{equation}
One can solve this recurrence using the generating function technique, from which one gets
\begin{subequations}
\begin{equation}
\label{nu-exact}
P(n,3n) =3^{-n}
\end{equation}
This simple solution can be verified by substitution of \eqref{nu-exact} into the recurrence \eqref{nu-n}.
 
A similar analysis gives two other series of probabilities of maximally sparse jammed states:
\begin{align}
\label{nu-exact-1}
&P(n,3n-1) =\frac{3n+2}{5}\,\frac{1}{3^{n-1}}\\
\label{nu-exact-2}
&P(n,3n-2) =\frac{63n^2+45n+8}{350}\,\frac{1}{3^{n-2}}
\end{align}
\end{subequations}
The exact results \eqref{nu-exact}--\eqref{nu-exact-2} are the more precise versions of the asymptotic \eqref{Prob:extreme-min} following from the cumulant generating function.

Thus, in the realm of the RSA model, the defect-free arrays arise with exponentially small probabilities. The defect-free arrays \eqref{Z3}--\eqref{Z2} of Rb atoms excited to Rydberg states have been recently assembled experimentally by Bernien {\em et al.} \cite{Lukin17}.

\subsection{Probability distribution $P(N,L)$}

The probability distribution $P(N,L)$ satisfies a recurrence relation
\begin{eqnarray}
\label{PNL:rec}
L P(N,L) \!=\! \sum_{k=1}^{L}\sum_{i+j=N-1}P(i,k-2)P(j,L-k-1)
\end{eqnarray}
which holds for all $L\geq 1$ if we set 
\begin{equation}
\label{PN0}
P(N,L)=\delta_{N,0} \qquad\text{for all} \quad L\leq 0. 
\end{equation}
One can verify that $P(N,1)=\delta_{N,1}$ and  $P(N,2)=\delta_{N,1}$. We now introduce the generating function
\begin{equation}
\label{Pxy}
\mathcal{P}(x,y)=\sum_{N\geq 0}\sum_{L\geq 0}P(N,L) x^N y^L
\end{equation}
This definition implies that
\begin{equation}
\label{LHS}
\sum_{N\geq 0}\sum_{L\geq 0} LP(N,L) x^N y^{L-1} = \frac{\partial \mathcal{P}}{\partial y}
\end{equation}
Using \eqref{PNL:rec}--\eqref{LHS} we obtain a differential equation for the generating function
\begin{equation}
\label{P-GF}
\frac{\partial \mathcal{P}}{\partial y} = x[1+y\mathcal{P}]^2
\end{equation}
which is mathematically identical to Eq.~\eqref{Phi:eq}. The solution of \eqref{P-GF} subject to the boundary condition $\mathcal{P}(x,0)=1$ is thus identical to \eqref{FLx} in different notation:
\begin{equation}
\label{Pxy:sol}
\mathcal{P}(x,y) = \frac{1+\sqrt{x} \tanh(y\sqrt{x})}{1-y+(\frac{1}{\sqrt{x}}-y\sqrt{x})\tanh(y\sqrt{x})}
\end{equation}

The occupation number $N$ always lies within the bounds
\begin{equation}
\label{bounds}
\left\lfloor \frac{L+2}{3}\right\rfloor \leq N \leq \left\lfloor \frac{L+1}{2}\right\rfloor
\end{equation}
For instance, $5\leq N\leq 7$ when $L=14$;  expanding \eqref{Pxy:sol} one finds the corresponding probabilities
\begin{equation*}
P(5,14)=\tfrac{17}{405}, ~~ P(6,14)=\tfrac{1854632}{3274425}, ~~ P(7,14)=\tfrac{1282348}{3274425}
\end{equation*}

\section{Arbitrary Blockage Radius}
\label{sec:LMb}

In this section, we analyze the model with an arbitrary blockage radius and show that some characteristics remain analytically tractable. (In the context of ultra-cold Rydberg atoms, models with $b>1$ are often relevant; see, e.g., \cite{Les12a,Sachdev}.)  

\subsection{The average number of occupied sites}

The average number of occupied sites obeys 
\begin{equation}
\label{ALb:rec}
A_L = \frac{1}{L}\sum_{k=1}^L \big(A_{k-b-1}+1+A_{L-k-b}\big)
\end{equation}
This recurrence is applicable for all $L\geq 1$ after setting $A_j=0$ for $j\leq 0$. Using the generating function \eqref{Ax:def} we convert \eqref{ALb:rec} into a differential equation
\begin{equation}
\label{Axb:eq}
\frac{d A}{dx} = \frac{2x^b}{1-x}\,A +(1-x)^{-2}
\end{equation}
Solving \eqref{Axb:eq} subject to $A(0)=0$ yields 
\begin{equation}
\label{Axb:sol}
A(x) = \frac{e^{-2L_b(x)}}{(1-x)^2}\int_0^x dy\,e^{2L_b(y)}
\end{equation}
where we shortly write
\begin{equation}
L_b(x) = \sum_{j=1}^b \frac{x^j}{j}
\end{equation}
Using \eqref{Axb:sol} we extract the density of occupied sites
\begin{equation}
\label{jammed-b}
\rho_b =e^{-2L_b(1)}\int_0^1 dy\,e^{2L_b(y)}
\end{equation} 
in the $L\to\infty$ limit. 

For the minimal model we recover $\rho_1=(1-e^{-2})/2$. The next jammed density $\rho_2$ also admits an explicit expression through standard special functions:
\begin{equation}
\label{jammed-2}
\rho_2 =\frac{\sqrt{\pi}}{2e^4}\left[\text{Erfi}(1)+\text{Erfi}(2)\right]=0.2745509877\ldots
\end{equation}
where $\text{Erfi}(z)=-\sqrt{-1}\,\text{Erf}(\sqrt{-1}z)$ is an error function. The following jammed densities are 
\begin{equation*}
\rho_3\approx 0.200973, \quad \rho_4\approx 0.158455, \quad \rho_5\approx 0.130772
\end{equation*}

The maximally dense and maximally sparse jammed states provide obvious upper and lower bounds on the jammed density:
\begin{equation}
\frac{1}{2b+1} < \rho_b < \frac{1}{b+1} 
\end{equation}
These bounds hint that the limit
\begin{equation}
\lim_{b\to \infty} b\rho_b = C
\end{equation}
exists and further suggest that $\frac{1}{2}<C<1$. One can compute $C$ using \eqref{jammed-b} and performing an asymptotic analysis. First we re-write \eqref{jammed-b} as
\begin{equation}
\label{jammed-b-new}
\rho_b  = \int_0^1 dy\,\exp\left[ - 2\sum_{j=1}^b \frac{1-y^j}{j}\right]
\end{equation}
We then write $y=1-v/b$ and transform \eqref{jammed-b-new} into
\begin{equation}
\label{jammed-b-2}
b\rho_b  = \int_0^b dv\,\exp\left[ - 2\sum_{j=1}^b \frac{1-\left(1-\frac{v}{b}\right)^j}{j}\right]
\end{equation}
In the $b\to\infty$ limit, we can use the asymptotic relation $\left(1-\frac{v}{b}\right)^j\to e^{-vj/b}$ and replace summation over $j$ by integration over $u=vj/b$. This leads to 
\begin{eqnarray*}
C =\int_0^\infty dv\,\exp\left[ - 2\int_0^v du\, \frac{1-e^{-u}}{u}\right]
\end{eqnarray*}
known as the R\'{e}nyi's parking constant \cite{R58}.

The efficiency of the coverage is measured by the ratio of the jamming density $\rho_b$ to the maximal possible density: $\theta_b=(b+1)\rho_b$. This quantity monotonically decays as $b$ increases and it approaches $C=\theta_\infty$. Here are a few numerical values
\begin{equation*}
\begin{split}
\theta_1  & = 0.86466471676338\ldots\\
\theta_2  & = 0.82365296317734\ldots\\
\theta_3  & = 0.80389347991537\ldots\\
\theta_4  & = 0.79227591371305\ldots\\
\theta_5  & = 0.78463015586503\ldots\\
\theta_\infty & =  0.74759791502876\ldots    
\end{split}
\end{equation*}

\subsection{Full counting statistics and probability distribution $P(N,L)$}

The function $F(\lambda, L)$ satisfies the recurrence
\begin{equation}
\label{FLb:rec}
F(\lambda, L) = \frac{e^\lambda}{L}\sum_{k=1}^L F(\lambda, k-b-1) F(\lambda, L-k-b)
\end{equation}
which we recast into a differential equation for the generating function $\Phi=\Phi(\lambda, x)$ defined by 
Eq.~\eqref{Phi:def}: 
\begin{equation}
\label{Phi:b-eq}
\frac{d \Phi}{d x} = e^\lambda \left[\frac{1-x^b}{1-x}+x^b\Phi\right]^2
\end{equation}

This Riccati equation \eqref{Phi:b-eq}  appears exactly solvable only for the minimal $b=1$ model. It is still possible to determine the variance by analyzing $\partial_\lambda \Phi(\lambda,x)$  and $\partial_\lambda^2 \Phi(\lambda,x)$  at $\lambda=0$. These functions satisfy equations which can be deduced from Eq.~\eqref{Phi:b-eq}; by solving these equations one can determine the variance as we show in the appendix. 

We also note that the probability distribution $P(N,L)$ satisfies a recurrence relation
\begin{equation}
\label{PNL:rec-b}
L P(N,L) \!=\! \sum_{k=1}^{L}\sum_{i+j=N-1}P(i,k-b-1)P(j,L-k-b)
\end{equation}
generalizing \eqref{PNL:rec}. Using the generating function \eqref{Pxy} we recast the recurrence \eqref{PNL:rec-b} into a Riccati equation
\begin{equation}
\label{P-GF-b}
\frac{\partial \mathcal{P}}{\partial y} = x\left[\frac{1-y^b}{1-y}+y^b\mathcal{P}\right]^2
\end{equation}
which is mathematically identical to \eqref{Phi:b-eq} and appears unsolvable when $b=2,3,4,\ldots$. 

\subsection{Extremal jammed states}

The probabilities of the largest possible deviations, i.e. extremal jammed states, can be probed analytically using the same approach  as in Sec.~\ref{subsec:Extremal}. We start with maximally dense jammed states and choose for concreteness such states with $N_\text{max}=n$ in the chains of length $L=(b+1)n-b$. This 
\begin{equation}
\bullet\,\circ\,\circ\,\bullet\,\circ\,\circ\,\bullet\,\circ\,\circ\,\bullet\,\circ\,\circ\,\bullet\,\circ\,\circ\,\,\bullet
\end{equation}
is an example of such maximally dense jammed state with $N_\text{max}=6$ and $L=16$ for the model with the blockage radius is $b=2$. The probabilities $\mu_n=P(n,(b+1)n-b)$ satisfy the recurrence
\begin{equation}
\label{mu-n-b}
\mu_n = \frac{1}{(b+1)n-b}\sum_{k=0}^{n-1}\mu_k\mu_{n-k-1}
\end{equation}
and the boundary condition \eqref{mu-n:BC}. The generating function \eqref{mu:GF} satisfies a Riccati equation 
\begin{equation}
\label{mu-b:Ric}
(b+1)\,\frac{d\mu}{dx}=b\,\frac{\mu-1}{x}+\mu^2
\end{equation}
Solving this Riccati equation subject to $\mu(0)=1$ yields 
\begin{equation}
\label{mu-b:sol}
\mu= -\frac{\sqrt{bx}\left[I_{\beta-1}(z)+I_{\beta-3}(z)\right]+(1+2b)I_{\beta-2}(z)}
{2x\,I_{\beta-2}}
\end{equation}
Here $I_p$ is the modified Bessel function and we have used shorthand notation $\beta=(1+b)^{-1}$ and  $z=2\beta\sqrt{bx}$. 

\begin{figure}[ht]
\begin{center}
\includegraphics[width=0.4\textwidth]{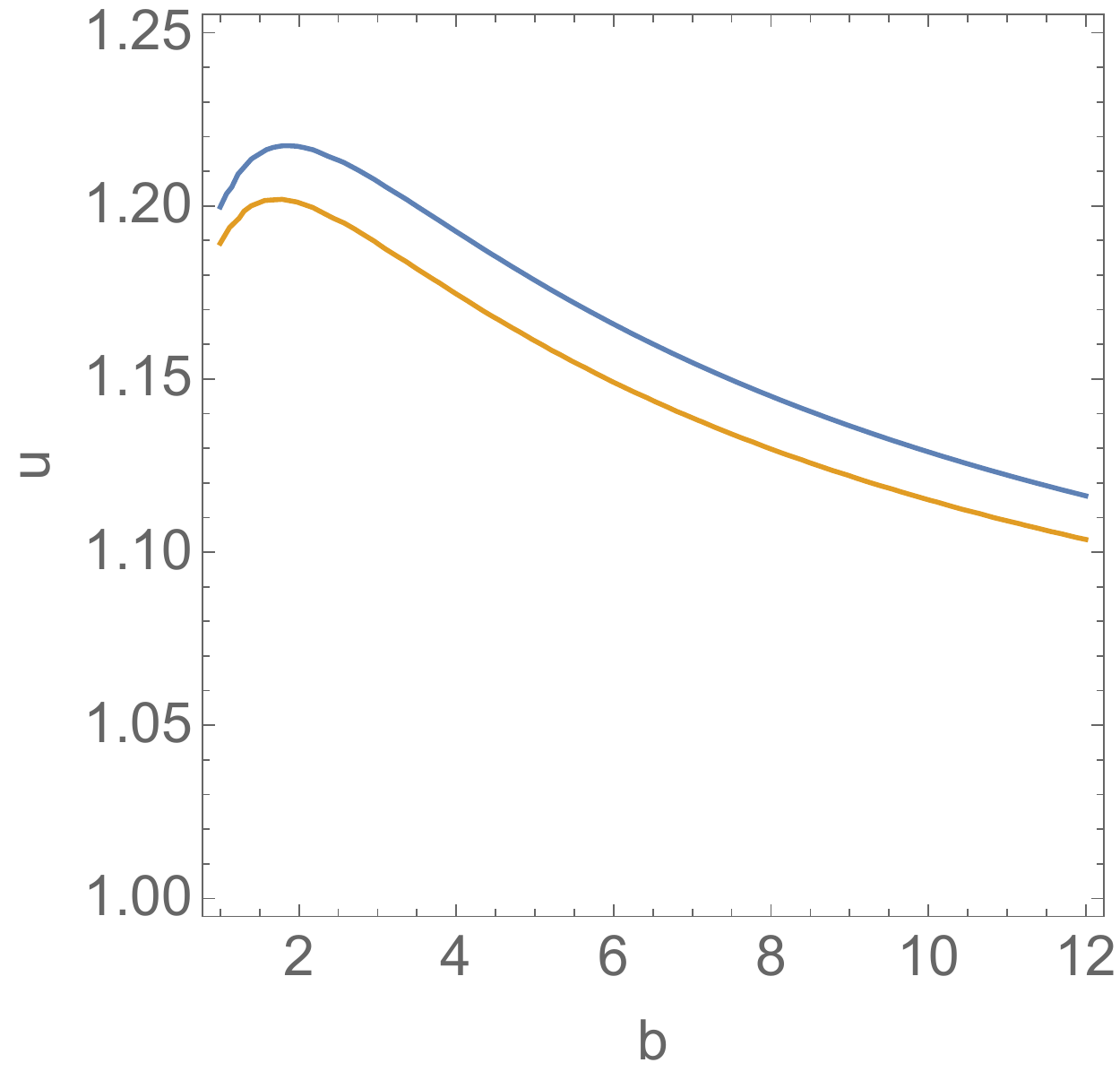}
\caption{Top curve: The exact value of $u(b)$ that determines the asymptotic decay, Eq.~\eqref{mu-y}, of the probability to reach a maximally dense jammed state. Bottom curve: The approximate value given by Eq.~\eqref{ub-approx}.}
\label{fig:ub}
  \end{center}
\end{figure}

The denominator in \eqref{mu-b:sol} vanishes when $x\to y$, where $y=y(b)$ is found from 
\begin{equation}
\label{root-b}
I_{\beta-2}\left(2\beta\sqrt{by}\right) = 0, \qquad \beta=(1+b)^{-1}
\end{equation}
This zero is simple and hence $\mu_n  \sim [y(b)]^{-n}$. Re-writing in terms of the length of the chain, we arrive at 
\begin{equation}
\label{mu-y}
 P(N_\text{max},L)\sim [u(b)]^{-L}, \qquad u(b)=[y(b)]^{1/(b+1)}
\end{equation}
The plot of $u(b)$ is shown in Fig.~\ref{fig:ub}. Only positive integer values of the blockage radius, $b\in \mathbb{Z}_+$, matter. A few first numerical values are
\begin{equation*}
\begin{split}
u(1) & = 1.19967864026\ldots\\
u(2) & = 1.21712361233\ldots\\
u(3) & = 1.20704834272\ldots\\
u(4) & = 1.19255384190\ldots\\
u(5) & = 1.17847101637\ldots
\end{split}
\end{equation*}

To extract more explicit results in the $b\to\infty$ limit we recall the series representation 
\begin{equation}
\label{I-beta}
I_{\beta-2}(z) = \sum_{m\geq 0}\frac{1}{m!\,\Gamma(m+\beta-1)}\,\left(\frac{z}{2}\right)^{2m-2+\beta}
\end{equation}
of the modified Bessel function. Since $z=2\beta\sqrt{by(b)}\ll 1$ when $b\gg 1$, i.e., $\beta\ll 1$, it suffices to keep the first three terms in \eqref{I-beta}. Using additionally $\Gamma(\beta)\simeq \beta^{-1}$ we obtain 
\begin{equation}
I_{\beta-2}(z) \simeq -\beta\left(\frac{z}{2}\right)^{-2}+\frac{1}{4}\,\left(\frac{z}{2}\right)^{2}
\end{equation}
in the leading order, from which we deduce
\begin{equation}
\label{ub-approx}
u\simeq \frac{\ln(b+1)}{2(b+1)}
\end{equation}
This is asymptotically exact when $b\to\infty$, but provides a reasonable approximation already for small $b$, see Fig.~\ref{fig:ub}.

Maximally sparse jammed states are easier to analyze than maximally dense jammed states. For instance, for chains of length $L=(2b+1)n$, we have $N_\text{min}=n$ and the probabilities $\nu_n=P(n,(2b+1)n)$ to generate such states satisfy the recurrence
\begin{equation}
\label{nu-n-b}
\nu_n = \frac{1}{(2b+1)n}\sum_{k=0}^{n-1}\nu_k\nu_{n-k-1}
\end{equation}
from which $\nu_n = (2b+1)^{-n}$.

\section{Conclusions}
\label{sec:D}

We have investigated a class of one-dimensional random sequential adsorption (RSA) models parametrized by an integer $b\geq 1$, the blockage radius defined as the number of sites on the left and right of an occupied site which must be empty. For this class of models,  the average number of occupied sites, the average densities near the boundaries, and the extreme probabilities to fall into maximally dense or minimally dense jammed states are all computable. We have mostly focused on large deviations and derived a  Riccati equation for the cumulant generating function. This Riccati equation is solvable only for the minimal model, $b=1$. When $b\geq 2$, it is in principle possible to extract cumulants consecutively via perturbation analysis. In the appendix, we have computed the variance. The complexity of such calculations quickly increases with the order of the cumulant, and the computation of the variance is already rather involved. The probabilities of maximally sparse jammed states exhibit a simple dependence on $b$; in the most clean situation of chains of length $L=(2b+1)n$ when $N_\text{min}=n$, the probabilities are $(2b+1)^{-n}$. The probabilities of maximally dense jammed states are more challenging: To extract the asymptotic behavior one must solve a Riccati equation for the generating function encoding these probabilities. We have done it for arbitrary $b$. 

The RSA models are tractable on some simple quasi-one-dimensional structures, e.g., on ladders. In particular, the RSA model with $b=1$ has been studied, and the jamming density has been computed \cite{Kutasov,Percus}.  The derivation \cite{Kutasov,Percus} of the jamming density is significantly more convoluted than in the purely one-dimensional setting, but perhaps the extreme probabilities would be easier to derive. The maximal density is again $\rho_\text{max}=\frac{1}{2}$, while the minimal density on the ladder, $\rho_\text{min}=\frac{1}{4}$, is smaller than in the one-dimensional setting. 

The RSA models often admit an analytical treatment on lattices without loops, e.g., on the Bethe lattices \cite{Evans} and some classes of infinite trees \cite{Fleurke}. Large tree-like graphs, i.e., graphs with few loops, can be also tractable. For instance, the RSA model on the sparse Erd\H{o}s-R\'enyi random graphs with $b=1$ has been studied, see \cite{ER:jamming,Dutch15,Dutch16,Dutch17}. It would be interesting to determine the full counting statistics of the occupation number for this RSA model.  

The occupation number is the simplest global quantity characterizing jammed states. More detailed global quantities are the final numbers of segments of different length. For the minimal model, there are short ($\bullet\,\circ\,\bullet$) and long ($\bullet\,\circ\,\circ\,\,\bullet$) segments in the final jammed state. The numbers of short and long segments, $N_s$ and $N_\ell$, are correlated random quantities;  it would be interesting to explore their joint distribution.  

We have analyzed  the statistics of the occupation number  of the jammed states. The RSA is a dynamical process, however, so it would be interesting to go beyond the jammed states and study the evolution of the occupation number $N(t)$, e.g., to probe the cumulants $\langle N^p(t)\rangle_c$. 

\vskip 1cm
\noindent 
{\bf Acknowledgments.} I am grateful to Chris Laumann for discussions.

\appendix*
\section{Calculation of the variance}
\label{ap:var}

The Riccati equation \eqref{Phi:b-eq}  seems solvable only for the minimal model, $b=1$. One can circumvent solving \eqref{Phi:b-eq} by recalling that we need only compute the derivatives of the rate function $U(\lambda)$ at $\lambda=0$. To compute the average and the variance, it suffices to determine the first and the second derivatives. Below we extract these derivatives from the functions
\begin{equation}
A(x)=\partial_\lambda \Phi(\lambda,x)\big|_{\lambda=0}\,,  \quad
B(x)=\partial_\lambda^2 \Phi(\lambda,x)\big|_{\lambda=0}
\end{equation}
Since $\Phi(\lambda,0)=1$, we have
\begin{equation}
\label{AB:BC}
A(0)=B(0)=0
\end{equation}
Below we also use 
\begin{equation}
\label{Phi:0}
\Phi(0,x) = \frac{1}{1-x}
\end{equation}
which follows from $F(0,L)\equiv 1$, see \eqref{F:def}, and the definition \eqref{Phi:eq} of the generating function $\Phi(\lambda,x)$. 

Differentiating Eq.~\eqref{Phi:b-eq} over $\lambda$, setting $\lambda=0$ and using  \eqref{Phi:0} we find that $A(x)$ satisfies 
Eq.~\eqref{Axb:eq}. The boundary condition is also the same, $A(0)=0$, and therefore $A(x)$ is given by \eqref{Axb:sol}. 

Differentiating Eq.~\eqref{Phi:b-eq} twice over $\lambda$, setting $\lambda=0$ and using  \eqref{Phi:0} we find that $B(x)$ obeys
\begin{equation}
\label{Bx:eq}
\frac{d B(x)}{dx} - \frac{2x^b}{1-x}\,B(x) = \frac{\mathcal{A}(x)}{(1-x)^2}
\end{equation}
where we have used the shorthand notation  
\begin{equation}
\label{A-cal}
\mathcal{A}(x)  =  1+4(1-x)x^{b}A(x)+2(1-x)^2 x^{2b}\,[A(x)]^2
\end{equation}
Integrating \eqref{Bx:eq} and using $B(0)=0$ we obtain 
\begin{equation}
\label{Bx:sol}
B(x) = \frac{e^{-2L_b(x)}}{(1-x)^2}\int_0^x dy\,e^{2L_b(y)}\,\mathcal{A}(y)
\end{equation}

We extract the first two cumulants by analyzing the singular behaviors of $A(x)$ and $B(x)$ near $x=1$.
Using Eq.~\eqref{Axb:eq} we find that $A(x)$ has a pole of degree two 
\begin{subequations}
\begin{equation}
\label{Ax:sing}
A(x) = \frac{\rho_b}{(1-x)^2} + \frac{2b\rho_b-1}{1-x}+O(1) 
\end{equation}
Similarly using \eqref{A-cal}--\eqref{Ax:sing} one finds that $B(x)$ has a pole of degree three
\begin{equation}
\label{Bx:sing}
B(x) = \frac{2\rho_b^2}{(1-x)^3}+ \frac{B_2}{(1-x)^2} + \frac{B_1}{1-x}+O(1) 
\end{equation}
\end{subequations}

Extracting the amplitudes from the exact solution \eqref{Bx:sol} is not straightforward. One may try to insert \eqref{Bx:sing} into \eqref{Bx:eq}, with the right-hand side computed with the help of \eqref{Ax:sing}. This gives $B_3=2\rho_b^2$, but does not fix $B_2$. To determine $B_2$ we combine \eqref{Bx:sol}  and \eqref{Bx:sing} to find
\begin{equation}
\label{B2}
B_2 = \lim_{x\to 1}\left[\int_0^x dy\,e^{2L_b(y)-2L_b(x)}\,\mathcal{A}(y)-\frac{2\rho_b^2}{1-x}\right]
\end{equation}
Equivalently,  $B_2 +2\rho_b^2= \rho_b J_b$ with 
\begin{equation}
\label{Jb}
J_b =\lim_{x\to 1}\int_0^x dy\left[e^{2L_b(y)-2L_b(x)}\,\frac{\mathcal{A}(y)}{\rho_b}-\frac{2\rho_b}{(1-y)^2}\right]
\end{equation}

To appreciate that the singular behaviors of $A(x)$ and $B(x)$ near $x=1$ give the cumulants we start by recalling that the function $\Phi(\lambda,x)$ has a simple pole at $y(\lambda)$, so
\begin{equation}
\label{Phi-sing}
\Phi(\lambda,x) = \frac{Y(\lambda)}{y(\lambda)-x} + O(1)
\end{equation}
when $x\to y(\lambda)$, with 
\begin{equation}
\label{Cy:BC}
Y(0)=1, \quad y(0)=1
\end{equation}
as it follows e.g. from Eq.~\eqref{Phi:0}. The average and the variance are found from the cumulant generating function $U(\lambda)=-\ln y(\lambda)$ to yield
\begin{subequations}
\begin{align}
\label{av}
\lim_{L\to\infty}L^{-1}\langle N\rangle  &= -y'\\
\label{var}
\lim_{L\to\infty}L^{-1}\langle N^2\rangle_c  &= (y')^2-y''
\end{align}
\end{subequations}
where $(\cdots)' = \frac{d (\cdots)}{d\lambda}\big|_{\lambda=0}$. 

Differentiating Eq.~\eqref{Phi-sing} over $\lambda$, setting $\lambda=0$ and using  \eqref{Cy:BC} we find
\begin{subequations}
\begin{align}
\label{Ax:sing-Cy}
A(x) &= -\frac{y'}{(1-x)^2} + \frac{Y'}{1-x}+O(1)\\
\label{Bx:sing-Cy}
B(x) &=\frac{2(y')^2}{(1-x)^3}  -\frac{y'' + 2 Y'y'}{(1-x)^2} + \frac{Y''}{1-x}+O(1)
\end{align}
\end{subequations}

Comparing the most diverging terms in \eqref{Ax:sing} and \eqref{Ax:sing-Cy} we conclude that 
\begin{subequations}
\begin{equation}
\label{y-rho}
y'=-\rho_b
\end{equation}
and thus confirm [see Eq.~\eqref{av}] that $\rho_b$ is indeed the jammed density. Comparing the next  most diverging terms in \eqref{Ax:sing} and \eqref{Ax:sing-Cy} we get 
\begin{equation}
\label{Y-rho}
Y' = 2b\rho_b-1
\end{equation}
\end{subequations}
Comparing \eqref{Bx:sing} and \eqref{Bx:sing-Cy} and using \eqref{y-rho}--\eqref{Y-rho} we get
\begin{equation}
\label{yC2}
y''=2\rho_b(2b\rho_b-1)-B_2
\end{equation}
Collecting these results we arrive at the final formula 
\begin{eqnarray}
\label{Q:gen}
Q_b =J_b + 1 - (1+4b)\rho_b
\end{eqnarray}
for the Mandel $Q$ parameter. The jammed density $\rho_b$ admits an integral representation, Eq.~\eqref{jammed-b}, while $J_b$ given by \eqref{Jb} requires taking the limit of an integral. (The naive replacement, $x\to 1$, in \eqref{Jb} gives erroneous integral representation.) It is still possible to extract accurate results, e.g., one finds $Q_2\approx -0.9518$ indicating that the statistics of the occupation number is also strongly sub-Poissonian when $b=2$. 

One can similarly compute higher cumulants. For instance, to determine the third cumulant one deduces the governing equation for $\partial_\lambda^3 \Phi(\lambda,x)\big|_{\lambda=0}$ from \eqref{Phi:b-eq},  solves it and from the singular behavior at $x=1$ extracts $y'''$; the third cumulant is then $3y' y'' - 2(y')^3-y'''$.


\begin{thebibliography}{99}
  
\bibitem{Evans} 
    J.~W.~Evans, Rev.\ Mod.\ Phys.\ {\bf 65}, 1281 (1993).
 
\bibitem{TT} 
    J.~Talbot, G.~Tarjus, P.~R. Van Tassel, and P.~Viot, Colloids Surfaces A {\bf 165}, 287 (2000).

\bibitem{Tor}
     S. Torquato, 
     {\it Random Heterogeneous Materials: Microstructure and Macroscopic Properties} 
     (Springer-Verlag, New York, 2002).
     
\bibitem{Adam}
     Z. Adamczyk, 
     {\it Particles at Interfaces: Interactions, Deposition, Structure} 
     (Elsevier, Amsterdam, 2006).

\bibitem{Newby13}
     P. C. Bressloff and J. M. Newby, Rev. Mod. Phys. {\bf 85}, 135 (2013).

\bibitem{Elimelech02}
     J. Y. Chen, J. F. Klemic, and M. Elimelech, Nano Lett. {\bf 2}, 393 (2002).

\bibitem{Kuznar03}
     M. Elimelech, J. Y. Chen, and Z. A. Kuznar, Langmuir {\bf 19}, 6594 (2003).

\bibitem{Floro06}
    J. L. Graya, R. Hull, and J. A. Floro,  J. Appl. Phys. {\bf 100}, 084312 (2006).

\bibitem{Yaish13}
     A. Katsman, M. Beregovsky, and Y. E. Yaish, Nano Studies {\bf 8}, 139 (2013);
     A. Katsman, M. Beregovsky, and Y. E. Yaish, J. Appl. Phys. {\bf 113}, 084305 (2013). 

\bibitem{TS10}
     S. Torquato and F. H. Stillinger, Rev. Mod. Phys. {\bf 82}, 2633 (2010).  
       
\bibitem{F39} 
    P. J. Flory, J. Amer.\ Chem.\ Soc.\ {\bf 61}, 1518 (1939).

\bibitem{R58} 
      A. R\'enyi, Publ.\ Math.\ Inst. Hung.\ Acad.\ Sci.\ {\bf 3}, 109 (1958).
     
\bibitem{DCA07} 
    M. R. D'Orsogna, T. Chou, and T. Antal, J. Phys.\ A {\bf 40}, 5575 (2007).
  
\bibitem{KM10}   
    P. L. Krapivsky and K. Mallick,  J. Stat. Mech. P07007 (2010).

\bibitem{Privman92}
    V. Privman, Phys. Rev. Lett. {\bf 69}, 3686 (1992).

\bibitem{PK94}   
    P. L. Krapivsky,  J. Stat. Phys. {\bf 74}, 1211 (1994).

\bibitem{JML02}   
    G. De Smedt, C. Godr\`{e}che, and J. M. Luck,  Eur. Phys. J. B {\bf 27}, 363 (2020).

\bibitem{Exp:00}     
    D. Jaksch, J. I. Cirac, P. Zoller, S. L. Rolston, R. C. C\^{o}t\'{e}, and M. D. Lukin,
    Phys. Rev. Lett. {\bf 85}, 2208 (2000).
   
\bibitem{Exp:05}     
     T. Cubel Liebisch, A. Reinhard, P. R. Berman, and G. Raithel, Phys. Rev. Lett. {\bf 95}, 253002 (2005).
     
\bibitem{Exp:12}     
     M. Viteau, P. Huillery, M. G. Bason, N. Malossi, D. Ciampini, O. Morsch, E. Arimondo, D. Comparat, 
     and P. Pillet, Phys. Rev. Lett. {\bf 109}, 053002 (2012).

\bibitem{Exp:13} 
     C. S. Hofmann, G. G\"{u}nter, H. Schempp, M. Robert-de-Saint-Vincent, M. G\"{a}rttner, J. Evers, S. Whitlock, 
     and M. Weidem\"{u}ller, Phys. Rev. Lett.  {\bf 110}, 203601 (2013).

\bibitem{Exp:14}     
     N. Malossi, M. M. Valado, S. Scotto, P. Huillery, P. Pillet, D. Ciampini, E. Arimondo, 
     and O. Morsch, Phys. Rev. Lett. {\bf 113}, 023006 (2014).       

\bibitem{Rydberg:FCS}
     H. Schempp et al, Phys. Rev. Lett. {\bf 112}, 013002 (2014).

\bibitem{Dutch14}
    J. Sanders, R. van Bijnen, E. Vredenbregt, and S. Kokkelmans, Phys. Rev. Lett. {\bf 112}, 163001 (2014).

\bibitem{Lukin17}
      H. Bernien, S. Schwartz, A. Keesling, H. Levine, A. Omran,   H.  Pichler,   S.  Choi,   A.  S.  Zibrov,   M.  Endres,
      M.  Greiner,  V.  Vuleti\'{c},    and  M.  D.  Lukin,
      Nature {\bf 551}, 579 (2017).

\bibitem{Abanin17}
     C.  J.  Turner,  A.  A.  Michailidis,  D.  A.  Abanin,  M.  Serbyn,  and  Z.  Papi\'{c}, 
     Nat. Phys. {\bf 14}, 745 (2018); Phys. Rev. B {\bf 98}, 155134 (2018). 
     
\bibitem{Lukin18}
      W. W. Ho, S. Choi, H. Pichler, and  M.  D.  Lukin,   Phys.  Rev.  Lett. {\bf 122}, 040603 (2019). 
     
\bibitem{Chris18}
      V. Khemani, C. R. Laumann, and A. Chandran,  Phys. Rev. B {\bf 99}, 161101(R) (2019). 

\bibitem{book}   
     P. L. Krapivsky, S. Redner and E. Ben-Naim,  {\it  A Kinetic View of Statistical Physics} 
     (Cambridge University Press, Cambridge, UK, 2010).

\bibitem{Zagier}
    D. Zagier, in: {\em First European Congress of Mathematics}, Vol. II (Paris, 1992), vol. 120 of Progress of Mathematics,
    pp. 497--512 (Birkh\"{a}user, Basel, 1994).

\bibitem{JML}
    J. M. Luck and A. Mehta, Phys. Rev. E {\bf 92}, 052810 (2015). 

\bibitem{Mandel}
      L. Mandel, Optics Lett. {\bf 4}, 205 (1979).

\bibitem{Fano}
     U. Fano, Phys. Rev. {\bf 72}, 26 (1947).

\bibitem{Les12a}
     C. Ates and I. Lesanovsky, Phys. Rev. A {\bf 86}, 013408 (2012). 

\bibitem{Sachdev}
     R. Samajdar, W. W. Ho, H. Pichler, M. D. Lukin, and S. Sachdev, Phys. Rev. Lett.  {\bf 124},103601 (2020). 
     
\bibitem{Kutasov} 
     A. Baram and D. Kutasov, J. Phys. A {\bf 25}, L493 (1992).
 
\bibitem{Percus}
     Y. Fan and J. K. Percus,  J. Stat. Phys. {\bf 66}, 263 (1992).
 
\bibitem{Fleurke}
    H. G. Dehling, S. R. Fleurke, and C. K\"{u}lske, J. Stat. Phys. {\bf 133}, 151 (2008).

\bibitem{ER:jamming}
    C. McDiarmid,  Ann. Oper. Res. {\bf1} 183 (1984). 

\bibitem{Dutch15}
    J. Sanders, M. Jonckheere, and S. Kokkelmans, Phys. Rev. Lett. {\bf 115}, 043002 (2015).

\bibitem{Dutch16}
     S. Dhara, J. S. H. van Leeuwaarden, and D. Mukherjee, J. Stat. Phys. {\bf 164}, 1217 (2016).
 
\bibitem{Dutch17}
    P.  Bermolen, M. Jonckheere, and J. Sanders, J. Stat. Phys. {\bf 169}, 989 (2017).

\end{thebibliography}
\end{document}